\documentclass[twocolumn, prb, showpacs]{revtex4}
\usepackage{graphics}
\begin{document}
\def\zt{Zlatko Te\v sanovi\' c}
\title{\bf Low-Temperature Specific Heat of an Extreme Type-II Superconductor at High Magnetic
Fields}
\newcommand{\q}{\bf q}
\newcommand{\be}{\begin{equation}}
\newcommand{\ee}{\end{equation}}
 \author{Amanda L. Carr}
 \author{John J. Trafton}
 \author{Sa\v {s}a Dukan } 
 \email{sdukan@goucher.edu} 
 \affiliation{Department of Physics and Astronomy, Goucher College, Baltimore, MD 21204}
 \author{Zlatko Te\v sanovi\' c }
 \affiliation{Department of Physics and Astronomy, Johns Hopkins University, Baltimore, MD 21218} 

\begin{abstract}
~ We present a detailed study of the quasiparticle contribution to the low-temperature specific heat of 
an extreme type-II superconductor at high magnetic fields. Within a T-matrix approximation for the self-energies 
in the mixed state of a homogeneous
superconductor, the electronic specific heat is a linear function of temperature with a linear-$T$ coefficient $\gamma _s(H)$
being a nonlinear function of magnetic field $H$. In the range of magnetic fields $H\agt (0.15-0.2)H_{c2}$ where our
theory is applicable, the calculated $\gamma _s(H)$ closely resembles the experimental data for the
borocarbide superconductor YNi$_2$B$_2$C. 
\end{abstract}
\maketitle

In the last two decades and particularly since the discovery of 
high temperature superconductors (HTS) almost all of the 
superconducting systems that hold the greatest promise 
for practical application 
are of the extreme-type-II variety. 
These materials
are characterized by high transition temperatures ($T_c$), 
high upper critical fields ($H_{c2}$), and can be defined 
as materials in which the semiclassical $H_{c2}(0)$ 
in units of Tesla becomes comparable to, or even larger than, $T_c$ 
in units of Kelvin. 
In such systems, at low temperature and 
high magnetic fields near and around the semiclassical $H_{c2}(0)$, 
Landau level quantization of electronic
energies {\it within} the superconducting 
state is well defined and has to be included in the description of the
superconducting instability \cite{sasa1, akera}. 
Such a regime in which the Landau level structure is well 
resolved , {\it i.e.}
cyclotron energy $\hbar \omega_c \geq \Delta (T,H),T,\Gamma  $ 
(where $\Delta (T,H)$ is the BCS gap, and $\Gamma $ 
is the scattering rate due to disorder), 
represents a large portion of the $H-T$ phase diagram 
of an intrinsically extreme 
type-II superconductor. This region can extend 
down to magnetic fields as low as $\sim (0.2-0.5)H_{c2}(0)$ and temperatures
as high as $\sim 0.3T_{c0}$. In contrast, the size of the similar
region in an ordinary type-II superconductor (such as Nb) 
is expected to be negligible and confined to an immediate vicinity
of $H_{c2}(0)$.\cite{dhva} Inside this
high-field, low-temperature regime the 
superconducting state fundamentally differs 
from the familiar low-field mixed phase 
of the Abrikosov-Gorkov theory \cite{abrikosov1}, 
primarily by the appearance of {\it gapless} quasiparticle excitations at
the Fermi surface. These gapless excitations reflect a coherent quasiparticle 
propagation over many unit cells of 
a closely packed vortex lattice with 
fully overlapping vortex cores 
\cite{sasa2}. 

	The presence of such low-lying quasiparticle 
excitations makes an $s$-wave, ``conventional" superconductor in a 
high magnetic field somewhat similar to an 
anisotropic, ``unconventional" superconductor with 
nodes in the gap. In the low-temperature and 
high-field regime, however, the nodes in the gap reflect the 
{\em center-of-mass} motion of the Cooper 
pairs in the magnetic field, in contrast to  $d$-wave superconducting cuprates 
where such nodes are due to the {\em relative orbital} motion. 
This gapless behavior in three-dimensional systems is 
not restricted to fields very close to $H_{c2}$ 
but rather persists to a surprisingly low magnetic fields 
$H^*\sim (0.2-0.5)H_{c2}$.\cite{sasa2,pedro,nishio} 
Below $H^*$ gaps start opening up in the quasiparticle spectrum
and the system eventually reaches the low-field
regime of localized states in the cores of isolated, well-separated 
vortices \cite{norman}. Recently, an extensive 
numerical calculation of quasiparticle excitations in the mixed state for 
both $s$-wave and $d$-wave superconductors was performed \cite{kita} 
and it was found that for fields $H\agt 0.5H_{c2}$
no qualitative difference in behavior can 
be seen between $s$- and $d$-wave cases. They are both characterized by coherent
low-lying Landau level-like
quasiparticles excitations. However, a marked difference appears 
at lower fields $H\ll 0.5H_{c2}$, where an $s$-wave 
superconductor is clearly in the regime of localized, bound vortex core states
while a $d$-wave system 
still exhibits the extended nature of low-lying
quasiparticle excitations as predicted by 
Franz and Te\v{s}anovi\'{c} \cite{franz}.
Furthermore, it was shown that the high-field 
gapless character of the excitation spectrum is not destroyed 
by a moderate level of nonmagnetic impurities  present in either dirty 
homogeneous superconductor \cite{sasa3} 
or dirty inhomogeneous superconducting systems \cite{pedro1}.

The strongest evidence for Landau level 
quantization {\it within} the superconducting state comes from the 
experimental observation of de Haas-van Alphen 
(dHvA) oscillations in various A-15 and  borocarbide superconductors 
\cite{dhva}. The persistence of the dHvA 
signal deep within the mixed state of these three-dimensional
extreme type-II systems can be
attributed to the presence of a small portion 
of the Fermi surface containing gapless quasiparticle excitations, 
surrounded by regions where the gap is large \cite{sasa4,sasa5,maniv}. 
At the same time, careful measurements of thermal properties
({\it i.e.} thermal transport and/or specific heat) at low
temperatures and high magnetic fields are also expected 
to reveal the novel gapless behavior in these systems. The presence of
extended gapless quasiparticle states at low temperatures 
should lead to qualitatively different thermal behavior
than those found in an $s$-wave superconductor at low fields, 
where the number of quasiparticles excited above the gap is 
exponentially small and the only contribution might 
come from the bound states localized in the vortex cores. Recently, 
we studied the quasiparticle contribution to the thermal conductivities 
$\kappa _{ij}(\Omega ,T)$ of an extreme type-II
superconductor placed
in magnetic field $H$ such that $H_{c1}<<H\leq H_{c2}$. 
We examined the transport coefficients $\kappa _{ij}/T$ in the 
limit of $\Omega \rightarrow 0$
and $T \rightarrow 0$ and found that there was considerable 
enhancement of thermal transport in the mixed state of an $s$-wave 
superconductor due to the creation of 
gapless excitations in the magnetic field. This is in contrast to the zero 
field thermal transport which is exponentially 
small for an $s$-wave superconductor with no nodes in the gap.\cite{sasa6} 
The agreement of our theoretical
curves with the experimental data for the 
borocarbide superconductor LuNi$_2$B$_2$C  and A-15 superconductor 
V$_3$Si 
by Boaknin {\it et al.} \cite{boaknin,boaknin2}
is very good over a wide range of fields used in the
experiments.

The low-temperature electronic specific heat $C(T,H)$ is yet 
another probe of the 
quasiparticle excitations in the mixed state of a superconductor. 
In a fully gapped $s$-wave superconductor at low magnetic fields
there is an exponentially small contribution to 
$C(T,H)$ at low temperatures and 
the only significant contribution to $C(T,H)$  comes from 
the quasiparticles localized near the vortex axis 
\cite{caroli}. Assuming that the vortex core can be approximated
as a ``normal'' metal embedded in a superconducting medium, 
this contribution is then proportional 
to the quasiparticle density of states which is finite and approximately
equal to its normal state value. From here it follows that $C(T,H)$ 
varies linearly with $T$ and  
$\gamma _s(H)\equiv C(T,H)/T$ is proportional to
$H$ as $T\rightarrow 0$.
On the other hand, it was predicted that 
in unconvential $d$-wave superconductors the 
density of states and therefore the 
linear-$T$ specific heat coefficient 
varies as $\sqrt{H}$ at low fields $H\agt H_{c1}$. 
This field dependence is a consequence of 
the delocalized quasiparticles that can move 
along the nodal directions of the order parameter \cite{volovik}. 
Experiments
on HTS materials YBa$_2$Cu$_3$O$_{7-\delta}$ 
(Ref. 20) and La$_{2-x}$Sr$_x$CuO$_4$ (Ref. 21) 
have confirmed this 
theoretical prediction. The consensus has been reached  
that nonlinear field dependence
of $\gamma _s(H)$ in HTS systems is one of the 
signatures of an order parameter with d$_{x^2-y^2}$ symmetry. 
However, a number of 
experimental studies do not conform to this interpretation: 
a nonlinear $H$-dependence of $\gamma _s(H)$ in almost the entire
regime of the mixed state is observed in
$s$-wave superconductors, such as A-15's V$_3$Si (Ref. 22) and NbSe$_2$
(Refs. 23 and 24) as well as in borocarbides 
superconductors LuNi$_2$B$_2$C (Ref. 25) and YNi$_2$B$_2$C 
(Refs. 24 and 26). It is well established 
that A-15's are fully gapped 
superconductors at zero field while there might be
a significant anisotropy of the borocarbide's 
$s$-wave order parameter at very low 
fields $H\sim H_{c1}$. \cite{boaknin,izawa}
Sonier {\it et al.}\cite{sonier}
accounted quantitatively for this nonlinear 
behavior of $\gamma _s(H)$ in NbSe$_2$ 
by the expansion of the vortex cores
at low fields. On balance, it seems  that 
the specific heat behavior of $s$-wave 
superconductors in the whole regime of the mixed state is not fully
understood and merits further attention.

The purpose of this work is to examine in detail the 
quasiparticle contribution to the 
low-temperature specific heat in the 
mixed state of a three-dimensional $s$-wave extreme type-II 
superconductor starting from the high field limit. 
It was already suggested in Ref. 5 that 
in a pure superconducting system close 
to $H_{c2}$, the specific heat $C(H,T)$ should be an algebraic
function of temperature with the power dependent 
on the dimensionality of the system. 
This behavior was attributed to the strong dispersion 
around the gapless points in the quasiparticle excitation 
spectrum. In order to obtain analytical results, the authors of
Ref. 5 assumed a relatively small number $n_c$  of  Landau levels  
occupied by the electrons participating in 
the superconducting pairing ($n_{c}=E_F/\hbar\omega_c$, 
where $E_F$ is the Fermi energy and $\omega_c=eH/m*$ is the 
cyclotron frequency). However, 
this assumption is not expected to be quantitatively 
valid in the typical range of fields used in 
experiments \cite{ramirez, sanchez, nohara, nohara1, izawa}. 
On the contrary, the number of occupied 
Landau levels $n_c$ in the mixed
state $H_{c1}<<H\leq H_{c2}$ is often quite
large, typically $n_c\sim 30-270$ for 
borocarbide and $n_c\sim 250-4500$ for A-15 superconductors. 
The intention of the present work is to numerically evaluate 
the quasiparticle specific heat in the 
mixed state starting from the high-field limit 
of the Landau level pairing scheme, but 
under more realistic assumptions for the 
microscopic properties of materials studied so 
that a direct comparison can be made with available 
experimental data.

We begin by considering the density of 
states ${\cal N}(\omega, H)$ in a {\it dirty} 
but {\it homogeneous} superconductor
in the presence of nonmagnetic impurities in 
high magnetic field. In such a superconductor, 
the coherence length is much 
longer than the effective range of the impurity 
potential, so that under these conditions 
the order parameter in the 
mixed state $\Delta({\bf r})$ is not 
substantially affected and still forms a 
perfect vortex lattice. We follow  
Green's function perturbative approach to 
impurity effects in high magnetic fields 
developed in our previous work
 \cite{sasa3}. Normal and anomalous Green's 
functions for a clean superconductor are expanded in terms of the 
 complete set of
eigenfunctions in a magnetic sublattice 
representation (MSR). In the Landau gauge
${\bf A}=H(-y,0,0)$, the eigenfunction 
$\phi _{k_z,{\bf q},n} ({\bf r})$ belonging to $n$th 
Landau level can be written as
\begin{eqnarray}
\phi _{k_{z},{\q} ,n}({\bf r})=\sqrt{\frac{b_y}{2^{n}n!\sqrt{\pi }lV}} 
\exp {(ik_z\zeta )}
\nonumber\\
 \times\sum_{k}\exp {(i\frac{\pi b_x}{2a}k^2-ikq_yb_y)} 
\nonumber\\ 
\times \exp {[i(q_x+\frac{\pi k}{a})x-1/2(y/l+q_xl+\frac{\pi k}{a}l)^2]}
\nonumber\\
\times
H_{n}(\frac
{y}{l}+(q_x+\frac{\pi k}{a})l),
\label{phi}
\end{eqnarray}
where $\zeta $ is the spatial coordinate 
and $k_{z}$ is the momentum along the 
field direction, ${\bf a}=(a,0)$ and 
${\bf b}=(b_x,b_y)$ are the unit vectors 
of the triangular vortex lattice, $l=\sqrt{\hbar /eH}$ is the magnetic
length, and $V$ is the volume of the 
system. $H_{n}(x)$ is the Hermite polynomial of order $n$. 
Quasimomentum ${\q}$, perpendicular to the 
direction of the magnetic field, is restricted 
to the first magnetic brillouin zone (MBZ) spanned by vectors 
${\bf Q_1}=(b_y/l^2,-b_x/l^2)$ and 
${\bf Q_2}=(0,2a/l^2)$.\cite{sasa2} 
In this representation, the "Fourier transforms" of superconducting 
Green's functions in this quasimomentum
space expressed in the Nambu formalism can be written as 
\begin{eqnarray}
 \hat{{\cal G}}_n(k_z,{\bf q},i\omega)= 
 \frac{1}{(i\omega)^2-E_{n}(k_z,{\q})}
 \nonumber\\
 \times\left(\begin{array}{cc}
 i\omega+\epsilon_{n}(k_z) &-\Delta_{nn}({\q})\\
 -\Delta_{nn}^*({\q})& i\omega-\epsilon_{n}(k_z)
 \end{array}\right)
 \label{matrix}
 \end{eqnarray}
 where
\begin{eqnarray}
E_{n,p}(k_{z},{\q})=p\hbar\omega_c\pm \sqrt{\epsilon _{n}^{2}(k_z)+|\Delta _{n+p,n-p}({\q})|^{2}}
\nonumber\\
\epsilon _{n}(k_z)=\frac{\hbar ^2k_{z}^{2}}{2m}+\hbar \omega _{c}(n+1/2)-\mu
\label{epsilon}
\end{eqnarray}
is the quasiparticle excitation spectrum of the superconductor in high magnetic field near the
points $k_z=\pm k_{Fn}=\sqrt{2m(\mu-\hbar\omega_c(n+1/2))/\hbar^2}$. This spectrum is calculated  within the diagonal
 approximation \cite{sasa2,pedro}, where only the electrons belonging to mutually degenerate Landau
  Levels at the Fermi surface are involved in the superconducting pairing. Contributions to the pairing from the 
Landau levels separated by $\hbar\omega_c$ or more are included in the renormalization of the effective BCS coupling
constant $[g\rightarrow g(H,T)]$. \cite{sasa7} For quasiparticles near the Fermi surface 
 ($k_z\sim k_{Fn}$) it is enough to consider only the $E_{n,p=0}$ bands .   
 The gap function, $\Delta _{nm}({\q})$, in the MSR can 
  be written as
\begin{eqnarray}
\Delta _{nm}({\q} )=\frac{\Delta }{\sqrt{2}}\frac{
(-1)^{m}}{2^{n+m}\sqrt{n!m!}}\sum_kexp(i\pi \frac{b_x}{a}k^2)\times
\nonumber \\
exp(2ikq_yb_y-(q_x+\frac{\pi k}{a})^2l^2)
H_{n+m}[\sqrt{2}(q_x+\frac{\pi k}{a})l].
\label{delta}
\end{eqnarray} 
 The function $\Delta _{nm}({\q})$ turns to zero on the Fermi surface at the set of points in the 
MBZ with a strong linear 
dispersion in $q$. The excitations
from other, $p\neq 0$ in Eq. (\ref{epsilon}), bands are gapped by at least the cyclotron energy $\hbar \omega _c$ and their
 contribution to a superconductor's thermodynamics  can be neglected at low temperatures
  ($T\ll\Delta(T,H)<\hbar\omega_c$). Once the off-diagonal
 contribution is included in the superconducting pairing, the excitation spectrum cannot be written in the  
simple form, Eq. (\ref{epsilon}), and a closed analytic expression for the superconducting Green's function cannot be found.
A detailed study of the
effects of off-diagonal terms on the superconducting state has been pioneered by
Norman, MacDonald and collaborators (see Ref. 30 and references therein). Still, when these off-diagonal
 terms are treated analytically within the
perturbation theory of Ref. 6, the qualitative behavior of 
the quasiparticle excitations at the Fermi surface, as characterized by nodes in the MBZ, remains the same. 
This statement is correct in all orders of perturbation theory and therefore
is exact as long as the perturbative expansion itself is well defined, {\it i.e.} as long as the magnetic field is larger than
some critical field $H^*(T)$.  The critical field $H^*$ at $T\sim 0$ can
be estimated from the dHvA experiments to be $\sim 0.5H_{c2}$ for A-15 and $\sim 0.2H_{c2}$ for borocarbide
superconductors \cite{dhva}. Recent measurements of thermal transport in borocarbides in the mixed state \cite{boaknin}
suggest a 
strong anisotropy of the $s$-wave order parameter, so that the value for 
$H^*$ in these systems can be even lower than the estimate obtained from dHvA measurements.  
Once the magnetic field is lowered all the way to $H^*$, the
contribution of the off-diagonal pairing terms becomes essential \cite{norman1}
and gaps start
opening up at the Fermi surface signaling the crossover to the low-field
regime of quasiparticle states localized in the cores of widely separated vortices 
\cite{norman, kita, norman1}.

	In the presence of disorder the bare Green's 
function in Eq. (\ref{matrix}) is dressed via scattering through the diagonal
(normal) self-energy $\Sigma^N(i\omega )$ and off-diagonal (anomalous) self-energy 
$\Sigma^{A}_{nn}({\q},i\omega)$.\cite{sasa3} 
A dressed Green's function is obtained by replacing $\omega$ with $\tilde{\omega}$ and 
$\Delta_{nn}({\q})$ with $\tilde{\Delta}_{nn}({\q})$ in (\ref{matrix}) where
\begin{eqnarray}
i\tilde{\omega}\equiv i\omega - \Sigma ^N(i\omega)
\nonumber\\
\tilde{\Delta}_{nn}({\q})\equiv\Delta_{nn}({\q})+ \Sigma^{A}_{nn}({\q},i\omega).
\label{dressed}
\end{eqnarray} 
We follow a $T$-matrix approximation originally developed for heavy fermion superconductors \cite{hirch} and adapted 
by us to treat self-consistently
impurity scattering at high magnetic field \cite{sasa3}. Within this approximation 
both weak-scattering  and strong scattering limits can be treated on equal footing. However, we anticipate that the
 experimentally 
determined disorder 
parameters will put our calculation into the weak scattering limit of this theory with 
a dilute  concentration of impurities. Within a T-matrix approximation the self-energies $\Sigma ^N(i\omega)$ and
$\Sigma^{A}_{nn}({\q},i\omega)$ 
of the superconducting system are closely related to the diagonal (with respect to the magnetic translation group basis) 
$T$-matrix elements in a single-site approximation as \cite{sasa3}
\begin{eqnarray}
\Sigma ^N(i\omega )=n_i<T_{nn}^{11}(k_z,{\bf q},\omega )>_{{\bf R_i}} 
\nonumber\\
\Sigma _{nn}^A({\q};i\omega )=-n_i<T_{nn}^{12}(k_z,{\bf q},i\omega )>_{{\bf R_i}}
\label{definition}
\end{eqnarray}
where $<...>_{{\bf R_i}}$ denotes the average over the impurity positions and
$n_i$ is the impurity concentration. $T_{nn}^{ij}(k_z,{\bf q},i\omega )$ are the coefficients in the T-matrix expansion
over the complete set of MSR eigenstates. The $2\times 2$
matrix $\hat{{\cal T}}({\bf r},{\bf r'};i\omega )$ obeys the 
Lippmann-Schwinger equations
\begin{eqnarray}
\hat{{\cal T}}({\bf r},{\bf r'};i\omega )=U({\bf r})\delta({\bf r}-{\bf r'})\hat{\sigma }_z
+\int d{\bf r_1}U({\bf r})
\nonumber\\
\times \hat{\sigma }_z\hat{{\cal G}}({\bf r},{\bf r}_1;i\omega )\hat{\cal{T}}({\bf r}_1,{\bf r'};i\omega )
\label{lippman}
\end{eqnarray} 
where $\hat{{\cal G}}$-matrix elements are given by Eq. (\ref{matrix}) and $U({\bf r})=\sum_i U_0\delta ({\bf r}-{\bf R_i}) $ 
represents a short-range impurity potential with the location of impurities ${\bf R_i}$ taken to be randomly
distributed everywhere in the sample. The scalar scattering amplitude $U_0$ is assumed to be isotropic. In high magnetic
field we can assume that the scattering potential is weak compared to  the separation between Landau levels, given by
 $\hbar \omega _c$. Under these circumstances the electrons scatter into the states belonging to the same Landau level so that
 the off-diagonal (with respect to Landau level index $n$) $T$-matrix elements in Eq. (\ref{definition}) can be neglected. 
Note that this approximation might be
valid even under more general circumstances, when the effective scattering is
larger than $\hbar\omega_c$, as is the case in the Quantum Hall Effect (QHE) problems \cite{QHE}.
However, our case, unlike the QHE, also contains strong pairing potential and
thus the interference between the two is generally a rather formidable problem. Following the formalism outlined in Ref. 11,  Lippmann-Schwinger equations (\ref{lippman}) are averaged
 over the impurity position and from there can be solved as 
\begin{equation}
T_{nn}^{11}(i\omega )=
\frac{\left(b_y/V\sqrt{\pi }l\right)\sum _{k_z,{\bf k},m}
G_{m}^{11}(k_z,{\bf k},i\omega )}{1/U_0^2-\left[\left(b_y/V\sqrt{\pi }l\right)
\sum _{k_z,{\bf k},m}G_{m}^{11}(k_z,{\bf k},i\omega)\right]^2}
\label{ts3}
\end{equation}
and
\begin{eqnarray}
T^{21}_{nn}({\bf q};i\omega )=
-\left(\sqrt{2}b_y/V\sqrt{\pi }l\right)
f_{nn}^*({\q})
\nonumber\\
\times\frac{\sum _{k_z,{\bf k},m}f_{mm}({\bf k})G_{m}^{21}(k_z,{\bf k},i\omega)}
{1/U_0^2-\left[\left(b_y/V\sqrt{\pi }l\right)\sum _{k_z,{\bf k},m}
G_{m}^{11}(k_z,{\bf k},i\omega)\right]^2}
\label{ts4}
\end{eqnarray} 
where  $G_{m}^{ij}(k_z,{\bf k},i\omega)$ is a matrix element of a Nambu matrix (\ref{matrix}), $f_{nn}({\bf k})=
\Delta _{nn}({\bf k})/\Delta $ and $V$ is the volume of the system. $\sum_{{\bf k}}$ goes over the entire MBZ while
$\sum_{m}$ is over all the occupied Landau levels. Replacing $\omega $ with $\tilde {\omega}$ and 
$\Delta_{nn}({\q})$ with $\tilde{\Delta}_{nn}({\q})$ in (\ref{matrix}) with the help of definitions (\ref{dressed}) and
(\ref{definition}), Eqs. (\ref{ts3}) and (\ref{ts4}) can be brought to the form
\begin{widetext}
\begin{equation}
u=\frac{\omega }{\Delta }+\zeta \frac{\sum _{n=0}^{n_c}\left(m^*/4\pi ^3k_{Fn}N(0)\right)\int 
d{\q}(1-\sqrt{2}|f_{nn}({\q})|^2)u/\sqrt{u^2+|f_{nn}({\q})|^2}}
{c^2+\left[\sum_{n=0}^{n_c}\left(m^*/4\pi ^3k_{Fn}N(0)\right)\int d{\q}u/\sqrt{
u^2+|f_{nn}({\q})|^2}\right]^2} 
\label{unit}
\end{equation}
\end{widetext}
where $\zeta =\Gamma /\Delta $ and $u=\tilde{\omega }/\tilde{\Delta }$. $N(0)$ is the normal state density of states 
at the Fermi 
level. 
Disorder is 
characterized with two parameters: $\Gamma =n_i/N(0)\pi =(n_i/n)E_F$, 
which measures 
the concentration of impurities $n_i$ relative to the electron density
$n$, and $c=1/\pi N(0)U_0$, which measures 
the strength of the scattering potential. The weak-scattering limit in (\ref{unit}) is approached when $c^2$ is 
much larger than
the second term in the denominator of (\ref{unit}), typically when $c\sim 1$. The strong ({\it i.e.} unitary) scattering
limit is approached when $c=0$. The normal state inverse 
scattering rate $\Gamma _0 $ is found by taking $f_{nn}({\q})=0$
in (\ref{unit}) and letting $\omega \rightarrow 0$. This procedure yields 
\begin{equation}
\Gamma _0=\frac{\Gamma }{1+c^2}
\label{connect}
\end{equation}
and establishes the connection between the experimentally determined disorder
parameter $\Gamma _{0}$ and the parameters $\Gamma $ and $c$ in our 
theory.
Equation (\ref{unit}) is an implicit equation from which $u=f(\omega/\Delta)$ 
has to be calculated numerically. In fact, once the analytic continuation to real
frequencies  $i\omega =\omega +i\delta$ is performed, this equation transforms in a nonlinear system of
equations.
Finally, once $u$ is known the superconducting density of states and other thermodynamic 
quantities can be calculated.

	The superconducting density of states in the presence of disorder is defined as
\begin{equation}
{\cal N}(\omega, H)=
-\frac{1}{2\pi V}\Im m 
\sum _{n,k_z,{\q}}{\cal T}r\hat{{\cal G}}_n(k_z,{\q};i\tilde{\omega} )|_{i\omega =
\omega +i\delta }
\label{dens2}
\end{equation}
where $\hat{{\cal G}}_n(k_z,{\bf q},i\tilde{\omega})$ is a Nambu matrix (\ref{matrix}) in which the replacement
 $\omega $ with $\tilde {\omega}$ and 
$\Delta_{nn}({\q})$  with $\tilde{\Delta}_{nn}({\q})$ has been implemented. 
Once the analytic continuation to real frequencies is performed in Eq. (\ref{unit}) and Eq. (\ref{dens2}) and with
 the help of the definition $u={\tilde \omega}/{\tilde \Delta}$, 
the density of states (\ref{dens2}) can be obtained as
\begin{equation}
\frac{{\cal N}(\omega, H)}{N(0)}=\frac{1}{N(0)}\Im m\sum _{n=0}^{n_c}\frac{m}{4\pi ^3k_{Fn}}
\int d{\q}\frac{u}{\sqrt{|f_{nn}({\q})|^2-u^2}}.
\label{dens3}
\end{equation}
In deriving this equation, we assume that the impurity scattering does not change the ${\bf q}$-dependence  
of the quasiparticle excitation
spectrum,  {\it i.e.} ${\tilde \Delta}({\q})={\tilde \Delta}f_{nn}({\q})$ is given by Eq. (\ref{delta}). 
This assumption 
is shown to be valid for the high-field quasiparticle excitations close to the gapless points at the Fermi surface 
while it is less reliable for excitations gapped by large $\Delta$. \cite{sasa3} We are primarily interested in the
behavior of the specific heat when $T\rightarrow 0$ which is governed by the excitations around nodes while the 
contribution of the gapped regions in the MBZ is exponentially small. Therefore, this approximation seems a modest
 sacrifice in the quantitative accuracy when faced with the overwhelming numerical difficulty in determining fully 
 self-consistent ${\tilde \Delta}({\q})$ in the presence of the disorder.
 
	Once the superconducting density of states is obtained from Eq. (\ref{dens3}) the quasiparticle contribution to the 
specific heat in the mixed state at low temperatures can be 
computed as \cite{nicol}
\begin{equation}
C(T,H)=\frac{1}{2}\int _{-\infty}^{\infty}d\omega {\cal N}(\omega, H) \frac{\omega ^2}{T^2}\cosh ^{-2}\frac{\omega}{2T}
\label{specific}
\end{equation}
and then $\gamma _s(H)\equiv C(H)/T$ when $T\rightarrow 0$ can be determined. 
The most challenging part of our density of state and specific heat  
calculations for realistic superconducting materials in the range of the fields used in experiments
\cite{ramirez,sanchez,nohara,nohara1,izawa} is solving Eq. (\ref{unit}) when the number of Landau levels involved in the
superconducting pairing $n_c=E_F/\hbar\omega_c$ is very large. For the A-15 superconductor V$_3$Si, we
estimate $n_c\sim 241$ at $H_{c2}=18.5$ Tesla and $n_c\sim 4470$ at $H=1$ Tesla 
(we used an effective mass of $m^*=1.7m_e$ and Fermi velocity $v_F=2.8 \times 10^7$ cm/s from Ref. 34 in this 
estimate). For the borocarbide superconductor YNi$_2$B$_2$C, we get $n_c\sim 25$ at $H_{c2}=8$ Tesla (from Ref. 24)
and $n_c\sim 200$ at $H=1$ Tesla, obtained using $m^*=0.35m_e$, $v_F=2.5 \times 10^7$ cm/s, a
 normal scattering rate 
$\Gamma _0=0.53$ meV from the dHvA experiment of Terashima {\it et al.} \cite{dhva} (these values reproduce the mean
free path $l=1500 \AA$ reported for a clean sample by Nohara {\it et al.} \cite{nohara}). Faced with the overwhelming 
computational difficulty of solving a nonlinear system of equations that Eq. (\ref{unit}) becomes for $n_c\gg 100$ , 
we limit our study to the borocarbide superconductor YNi$_2$B$_2$C only. 
\begin{figure}
\scalebox{0.5}{\includegraphics{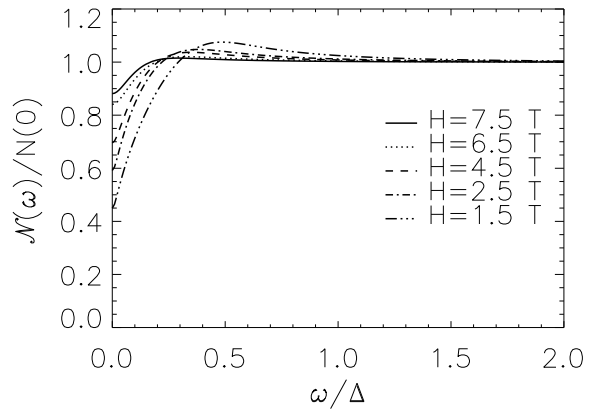}}
\caption{\label{fig1} Quasiparticle density of states ${\cal N}(\omega )$ in the mixed state rescaled by the 
normal state density of state $N(0)$ as a function of the reduced energy $\omega /\Delta$  computed from Eqs. (\ref{unit}),
(\ref{connect}) and (\ref{dens3}) for the borocarbide superconductor YNi$_2$B$_2$C when disorder parameter $c=0.65$. We have 
used experimentally determined values 
$H_{c2}=8$ Tesla and $T_c=15.4$ Kelvin 
from Ref. 24 as well as $m^*=0.35m_e$, $\Gamma _0=0.53$ meV and $\Delta =2.3$ meV from the dHvA measurement of Terashima
{\it et. al.} (Ref. 3).}
\end{figure}

	Figure 1 represents the quasiparticle density of
 states ${\cal N}(\omega )$ in the mixed state of the borocarbide superconductor YNi$_2$B$_2$C computed from
  Eq. (\ref{dens3})
 and rescaled by the 
normal state density of state $N(0)$ as a function of the reduced energy $\omega /\Delta$.
Disorder parameter $c$ in Eq. (\ref{unit}) is chosen to be $c=0.65$ which, at the same time,  determines a value for the 
second disorder parameter $\zeta$ 
to be $\zeta =0.33$ if the experimentally determined normal state
scattering rate of $\Gamma _0/\Delta$ from Ref. 3 is to be reproduced using Eq. (\ref{connect}). It can be seen in
Fig. \ref{fig1} that the density of states in the mixed state 
${\cal N}(\omega )$ at low energies diminishes as the magnetic field is lowered from $H\alt H_{c2}$ to $H\sim 0.2H_{c2}$.
 This is a consequence of the depletion of gapless or near gapless quasiparticle excitations at the Fermi surface 
 at lower magnetic fields. 
\begin{figure}
\scalebox{0.5}{\includegraphics{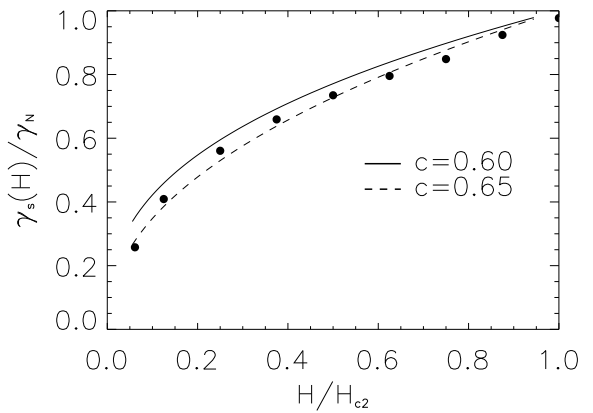}}
\caption{\label{fig2} The coefficient of the linear-T term in the quasiparticle specific heat at low temperature 
normalized by the Sommerfeld constant 
$\gamma (H)/\gamma _N$ for the borocarbide superconductor YNi$_2$B$_2$C as a function of reduced magnetic field $H/H_{c2}$.
 Full circles represent experimental data 
of Nohara {\it et al.}(Ref. 24), while lines represent our theoretical 
curves calculated from Eq. (\ref{specific})  with the help of Eqs. (\ref{unit}), (\ref{connect}) and (\ref{dens3}) 
for two values of the disorder 
parameter $c$ (see text).} 
\end{figure}

	The quasiparticle specific heat at low temperatures as computed from Eq. (\ref{specific}) is, to leading order, a linear
function of temperature $T$ due to the creation of a finite density of states at the Fermi level in Fig. 1. 
In Fig. 2, we plot $C(H,T)/T$
as $T\rightarrow 0$ ({\it i.e.} the coefficient of a linear-T term in the quasiparticle specific heat) normalized by the Sommerfeld 
constant $\gamma _{s}(H)/\gamma_N$ for YNi$_2$B$_2$C as a function of the reduced magnetic field $H/H_{c2}$.
Full circles represent the experimental data of Nohara {\it et al.} (Ref. 24).  
We plot two theoretical
curves calculated from Eq. (\ref{specific}) with $c=0.60$ (full line) and $c=0.65$ (broken line). The second disorder parameter
 $\zeta $ is calculated from
Eq. (\ref{connect}) where the experimentally determined  value for the normal state scattering rate $\Gamma _0=0.53$ meV 
from Ref.
3 is used. The values of the other physical quantities needed in our theory, effective mass $m^*=0.35$, 
BCS gap $\Delta =2.3$ meV and upper critical field $H_{c2}=8$ Tesla in $\Delta (H)=\Delta \sqrt{1-H/H_{c2}}$,
 are taken from experiments of Nohara {\it et al.} (Ref. 24) and/or
Terashima {\it et. al} (Ref. 3).
Fig. 2 demonstrates that the specific heat coefficient computed from the theory presented in this paper
exhibits a nonlinear dependence on the magnetic field $H$, {\it i.e.}, $\gamma _s(H)\sim H^{0.37}$ for $c=0.60$ and 
$\gamma _s(H)\sim H^{0.46}$ for $c=0.65$. This is in contrast to the linear $H$ dependence predicted for a 
fully gapped $s$-wave superconductor in the mixed state as $T\rightarrow 0$. We attribute this nonlinear behavior to the 
creation of coherent, gapless quasiparticle excitations at the Fermi surface in the mixed state of an extreme 
type-II superconductor
at high magnetic
field $H$ such that $H^*\alt H <H_{c2}$. The estimated critical field $H^*$ for pure 
YNi$_2$B$_2$C is $\sim (0.15-0.2)H_{c2}$. \cite{comment} However, this estimate depends on the value of the $s$-wave
gap function and can be much
 smaller than $0.15H_{c2}$ if the value of the minimum gap in the strongly anisotropic $s$-wave case is significantly different
from the accepted BCS value \cite{boaknin}. Furthermore, it seems that the unusual behavior of the specific heat coefficient
$\gamma _s(H)$ is also a consequence of disorder present in the superconducting system.  The finite amount of
impurities in the system leads to the creation of a finite 
density of states at the Fermi level ${\cal N}(0)$, in contrast to the perfectly clean superconductor where
 ${\cal N}(\omega)$ exhibits an algebraic behavior at low energies \cite{sasa3}. 
	
	 In summary, we have computed the  quasiparticle contribution to the specific heat $C(H,T)$ within the T-matrix 
 approximation 
 for the self-energies and found a nonlinear behavior of $\gamma _s(H)\equiv C(H,T)/T$ when $T\rightarrow 0$. 
 In the range of magnetic fields where
 our theory is applicable $H^*\leq H < H_{c2}$, the calculated $\gamma _s(H)$ closely resembles the experimental data for the 
 borocarbide superconductor  YNi$_2$B$_2$C.\cite{nohara}    

This work is supported by an award from Research Corporation (A.L.C., J.J.T. and S.D) and by NSF grant No. DMR00-94981
(Z.T.).


\begin{thebibliography}{99}
\bibitem{comment} In a clean s-wave superconductor $x=H^*/H_{c2}$ can be estimated as a solution of the equation 
$x^3=\frac{2\Delta^4}{\pi E_F(\hbar \omega _{c2})^3}(1-x)^2$ and yields $H^*\sim (0.15-0.2)H_{c2}$ for the system in question.
\end{thebibliography}
\end{document}